| | |
|---|---|
| **\*Title:** | Structural Characterization of an Ionic Liquid in bulk and in nano-confined environment from MD simulations |
| **\*Authors:** | Nataša Vučemilović-Alagić,[a,b] Radha D. Banhatti,[a] Robert Stepić,[a,b] Christian R. Wick,[a,b] Daniel Berger,[c] Mario Gaimann,[b] Andreas Bear,[b] Jens Harting,[c] David M. Smith\*,[a] and Ana-Sunčana Smith\*,[a,b] |
| **\*Affiliations:** | [a]Group for Computational Life Sciences, Department of Physical Chemistry, Ruđer Bošković Institute, Bijenička 54, 10000, Zagreb <br> [b]PULS Group, Institute for Theoretical Physics, IZNF, FAU Erlangen-Nürnberg, Cauerstraße 3, 91058, Erlangen <br> [c]Forschungszentrum Jülich GmbH, Helmholtz Institut Erlangen-Nürnberg, Fürther Straße 248, 90429, Nürnberg |
| **\*Contact email:** | \*E–mail: david.smith@irb.hr <br> \*E–mail: ana-suncana.smith@fau.de |
| **\*Co-authors**: | Nataša Vučemilović-Alagić- nvucemil@irb.hr <br> Radha D. Banhatti-Radha.Dilip.Banhatti@irb.hr <br> Robert Stepić - Robert.Stepic@irb.hr <br> Christian R. Wick - Christian.wick@fau.de <br> Mario U. Gaimann - Mario.gaimann@gmx.de <br> Andreas Baer - Andreas.baer@fau.de <br> Jens Harting - j.harting@fz-juelich.de <br> Daniel Berger - Daniel.x.berger@googlemail.com |
| **\*CATEGORY:** | <u>239</u>: Surfaces and Interfaces |

Data Article

Title: Structural Characterization of an Ionic Liquid in bulk and in nano-confined environment from MD simulations

Authors:

Nataša Vučemilović-Alagić,[a,b] Radha D. Banhatti,[a] Robert Stepić,[a,b] Christian R. Wick,[a,b] Daniel Berger,[c] Mario Gaimann,[b] Andreas Bear,[b] Jens Harting,[c] David M. Smith*[,a] and Ana-Sunčana Smith*[,a,b]

Affiliations:
[a]Group for Computational Life Sciences, Department of Physical Chemistry, Ruđer Bošković Institute, Bijenička 54, 10000, Zagreb
[b]PULS Group, Institute for Theoretical Physics, IZNF, FAU Erlangen-Nürnberg, Cauerstraße 3, 91058, Erlangen
[c]Forschungszentrum Jülich GmbH, Helmholtz Institut Erlangen-Nürnberg, Fürther Straße 248, 90429, Nürnberg

Contact email:
*E–mail: david.smith@irb.hr
*E–mail: ana-suncana.smith@fau.de

Abstract
This article contains data on structural characterization of the [C2Mim][NTf2] in bulk and in nano-confined environment obtained using MD simulations. These data supplement those presented in the paper "Insights from Molecular Dynamics Simulations on Structural Organization and Diffusive Dynamics of an Ionic Liquid at Solid and Vacuum Interfaces"[1], where force fields with three different charge methods and three charge scaling factors were used for the analysis of the IL in the bulk, at the interface with the vacuum and the IL film in the contact with a hydroxylated alumina surface. Here, we present details on the construction of the model systems in an extended detailed methods section. Furthermore, for best parametrization, structural and dynamic properties of IL in different environment are studied with certain features presented herein.

Keywords ionic liquid, nanoconfinement, molecular dynamics simulations, radial distributions, interface-normal number density

**Specifications Table**

| | |
|---|---|
| Subject area | Computational physical chemistry |
| More specific subject area | Molecular dynamics (MD) of ionic liquids in nano-confined environment |
| Type of data | Graph, figure |
| How data was acquired | NVT Molecular dynamics simulations at T = 300K via the Langevin algorithm, from which the first 100 ns were required for complete interfacial equilibration. Data processing and visualization using Gromacs analysis tools and self-developed scripts, as well as Xmgrace and molecular viewer VMD |
| Data format | *Raw and analyzed* |
| Simulation factors | *Brief description of set up and design of model systems* |
| Simulation features | *Classical MD; non-polarizable force field* |
| Data source location | *Group for Computational Life Sciences, Department of Physical Chemistry, Ruđer Bošković Institute, Bijenička 54, 10000, Zagreb* |
| Data accessibility | |
| Related research article | *N. Vučemilović-Alagić, R. D. Banhatti, R. Stepić, C. R. Wick, D. Berger, M. U. Gaimann, A. Baer, J. Harting, D. M. Smith and A-S. Smith, Insights from Molecular Dynamics Simulations on Structural Organization and Diffusive Dynamics of an Ionic Liquid at Solid and Vacuum Interfaces*, J. Colloid and Interface Science, **2019**, *submitted.* |

**Value of the Data**

- The simulated system is a prototype IL which is widely used in various technological applications.
- We explore the thickness of the stratified fluid at the solid and vacuum interfaces and determine the depth of the film necessary to recover bulk IL behavior in confinement.
- We determine the position dependent orientation of the cations and anions at the interfaces and hence provide an understanding of dominating interactions.
- The role of hydrogen bonding is shown to influence the arrangement at the solid-liquid interface as checkerboard of alternating anions and cations as evident both from INND plot and from MD snapshot, rather than a bilayer arrangement of IL at a charged solid interface.

**Data**

In this article we present structural data acquired in molecular dynamics simulations of an IL consisting of imidazolium based cations $[C_2Mim]^+$ and prototypical anions $[NTf_2]^-$ (Fig. 1a). Three different model systems are explored: a) the bulk IL (L), b) IL in between two vacuum slabs (V-L-V), and c) IL placed on a slab of sapphire surface (S-L-V)) and a 80 nm large slab of vacuum (Figs. 1b-d). The latter are today commonly used for Supported Ionic Liquid Phase

(SILP) catalysis[2,3,4] or Solid Catalysis with an Ionic Liquid Layer[5,6,] (SCILL).  In Fig. 2, we present the Interface Normal Number Densities of  the V-L-V and the S-L-V model systems which are used to identify a proper bulk region, shown as a pink slab, in every geometry so that the influence of one interface on the structural arrangement of the IL at the other interface is minimized. Fig. 3 shows that the total in-plane correlation functions ($h_{xy}$) is nearly the same for the bulk region of all three model systems.

Figures 4 to 7 contain data obtained for the L system. In Fig. 4, the radial distribution functions (rdfs) are shown with RESP-HF/0.9 parametrization, obtained from last 30 ns of the production runs of L system. A detailed comparison to the experimental structure factors (cf. Fig. S1 of Supporting Information to Ref. 1), demonstrates a good reproduction of the average structural features.  In Fig. 5 the rdfs of the H9 of cation (cf. Scheme 1 of Ref. 1) with nitrogen, sulphurs and oxygens of the anion, obtained for all three charge methods are compared for the scaling factor of 0.9. For details see Ref. 1. Fig. 6 shows results of our QM calculations using Gaussian09 software[7] that the preferred conformation of the anion in the gas phase is *trans*. In the L system, depending on the charge methods and charge scaling this conformation is *trans/gauche* or pure *cis* (cf. Fig. 2 of Ref.1).  Fig. 7 on the other hand shows that the distributions of the cation dihedral C2–N3–C7–C8 are practically independent of the chosen charge method and scaling factor.

Figures 8 to 12 describe various features of the structural organization of the IL at the solid and liquid interfaces, using RESP-HF/0.9 parametrization chosen as the optimal force field. Figure 8 presents structural ordering visible close to the solid-liquid interface from a snapshot from the MD simulations. INND profiles close to the solid liquid interface are presented in Fig. 9. Note that here and in further plots z = 0 is defined at the top of the sapphire surface. Figure 9 reveals the role played by hydrogen bonding between the hydroxylated sapphire surface and with the oxygens in the anion and with the hydrogen on the ring of the cation. This deduced orientation of the ring is confirmed by a probability distribution analysis presented in Fig. 10.  Both of these imply a checkerboard arrangement rather than the usually expected bilayer arrangement of the ions of the IL.[1] INND profiles presented in Fig. 11 for both S-L-V or a V-L-V model systems with sampling times of 160 and 100 ns respectively establish that that the behavior of the IL at the vacuum interface is the same in both systems. This enables an examination of the INND profiles at the vacuum interface as presented in Fig. 12, which shows that here there is no real possibility of bonding unlike at the solid interface. While the less polar $CF_3$ groups of the anion point towards the vacuum, it is the alkyl side chain of the cation that are closest to the vacuum interface. Detailed discussion of structural organization at the two interfaces and its implications for dynamics are discussed in detail in Ref. 1

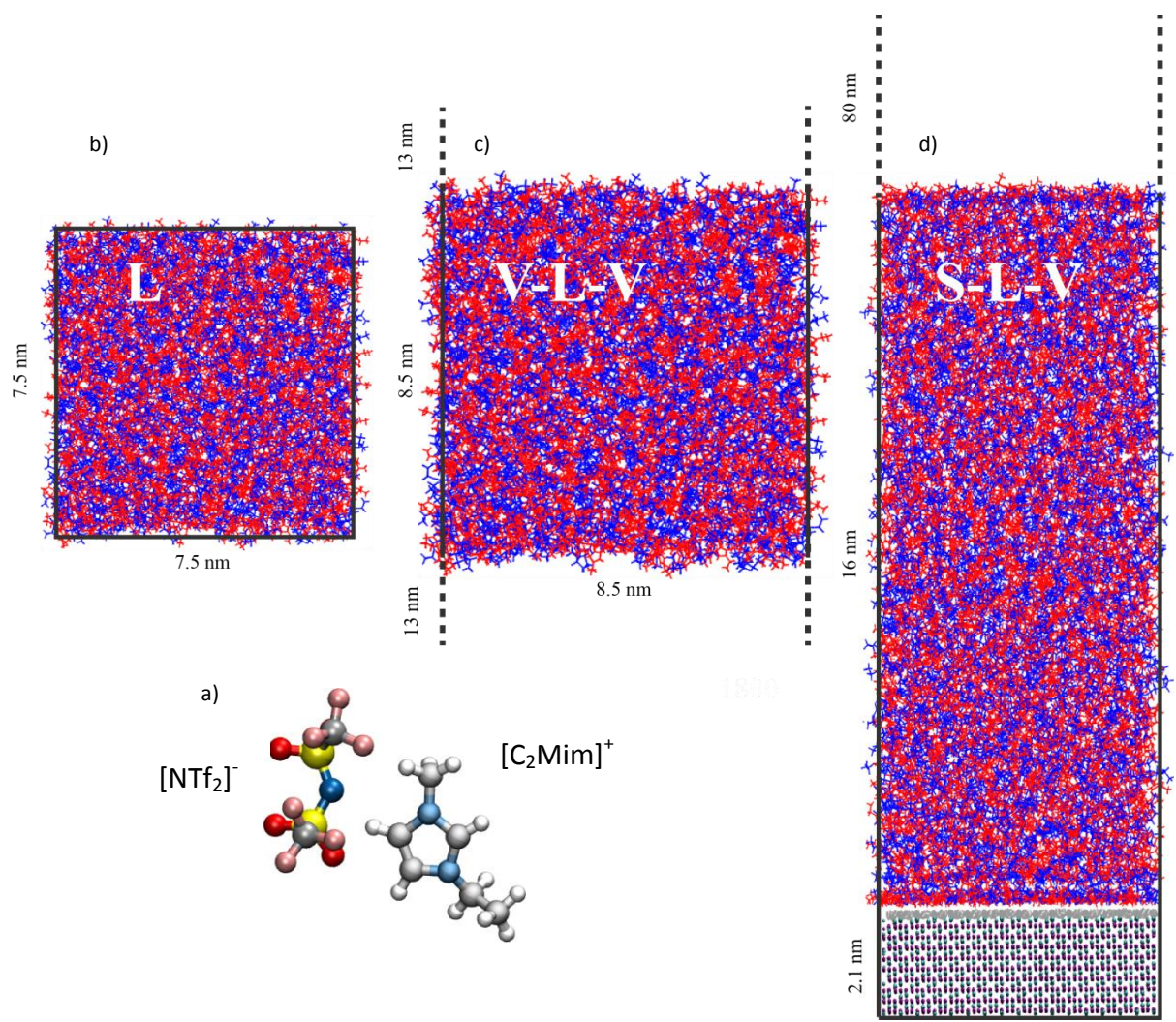

**Fig. 1.** a) The cation and anion constituting the ionic liquid (IL). b) The bulk liquid (L model system) consisting of 1000 ion pairs. Cations and anions are shown in red and blue respectively. c) The IL consisting of 1400 ion pairs between two vacuum slabs (V-L-V model system). d) The IL consisting of 1800 ion pairs in an S-L-V model system; the solid substrate is a fully hydroxylated (0001) slab of sapphire of dimension 7.57 nm x 6.29 nm x 2.12 nm.

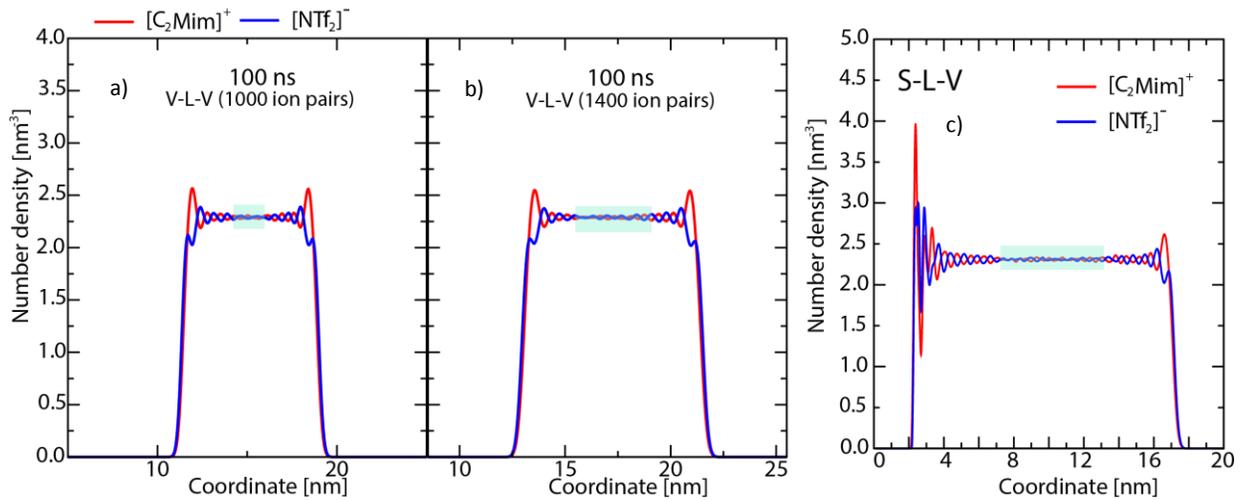

Fig 2. Interface Normal Number Density (INND): a) Non-homogenised bulk region of thickness 2.5 nm in V-L-V system with 1000 ion pairs; b) Larger and more homogenous bulk region (4nm) in V-L-V system containing 1400 ion pairs; c) S-L-V system with 1800 ion pairs – the homogenized bulk region is almost 7 nm. Note that here the choice of origin of the coordinate system is at the edge of the simulation box for both V-L-V and S-L-V systems. Only the relevant section is shown.

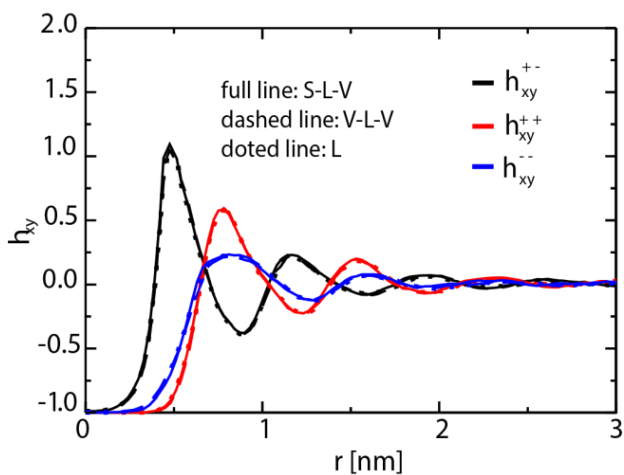

Figure 3. Total in-plane correlation functions ($h_{xy}$) in the bulk region of the L, L-V-L and S-L-V systems with sampling times of 70, 100 and 160 ns respectively.

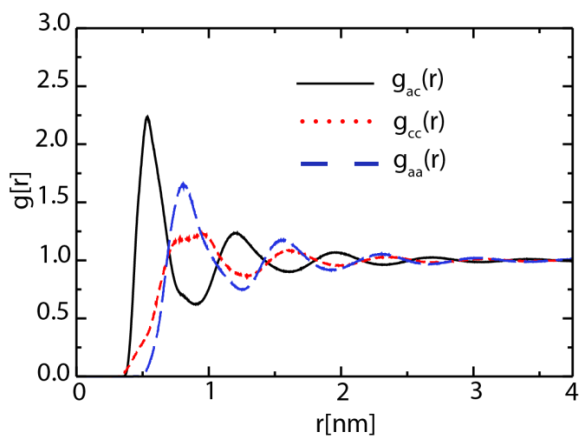

Figure 4. Radial distribution functions (rdfs), g(r), as function of distance, r, shown here for RESP-HF/0.9 parametrization. Black line: rdfs between centre of masses of cation [C2Mim]$^+$ and anion [NTf2]$^-$ ; red (blue) lines: rdfs between centre of masses of cations (anions). Rdfs were obtained with GROMACS tool using last 30 ns of the production runs of L system.

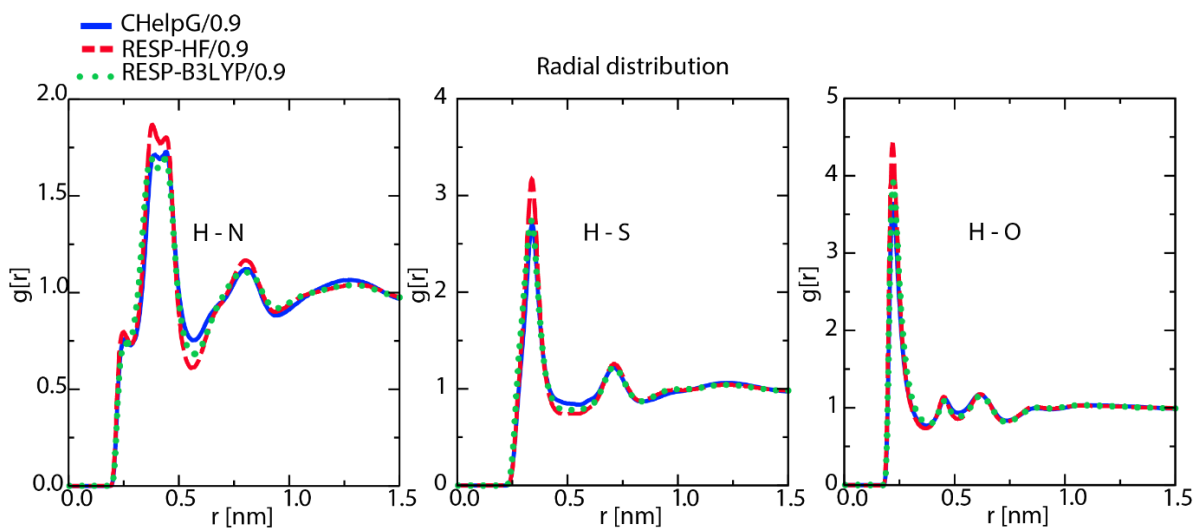

Figure 5. Radial distribution function of the H9 from cation with nitrogen, sulphurs and oxygens from anion obtained by three charge methods in the L system: blue (CHelpG/0.9), red (RESP-HF/0.9), green (RESP-B3LYP/0.9).

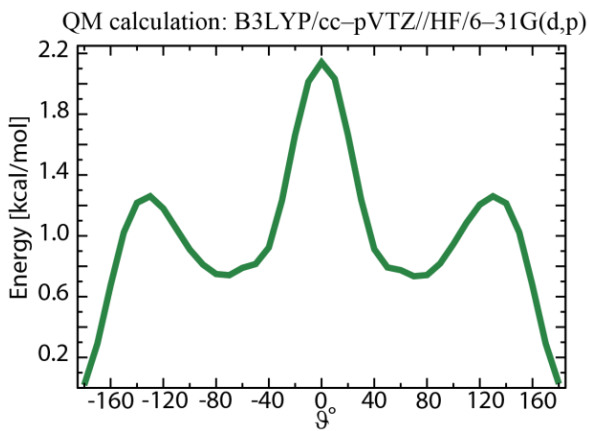

Figure 6. From quantum mechanical (QM) calculations at the B3LYP/cc–pVTZ//HF/6–31G(d,p) level of theory combined with an IEFPCM ($\varepsilon = 4.335$) continuum dielectric model mimicking solvent polarization, the anion can be seen to prefer a *trans* conformation. All QM calculations were performed using the Gaussian09 software package.[7]

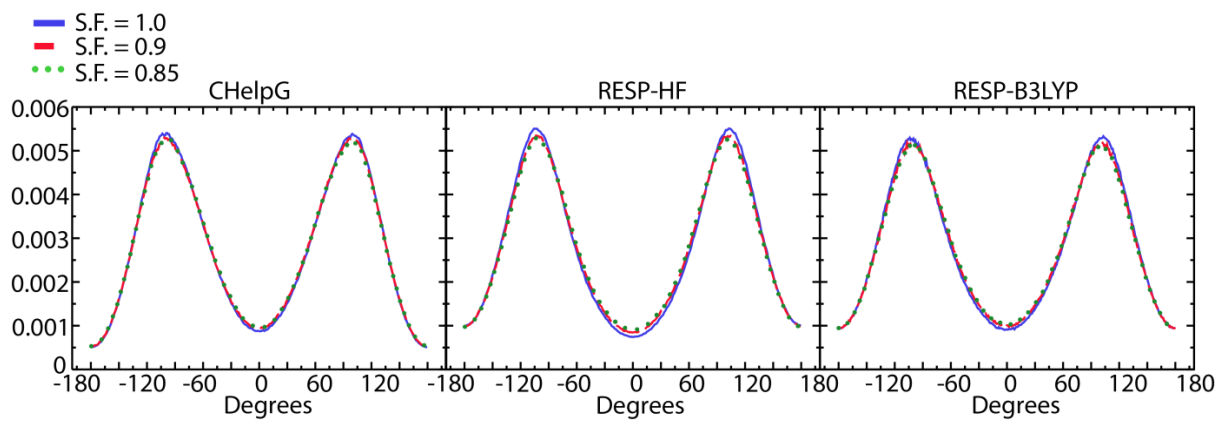

Figure 7. The distributions of the cation dihedral C2–N3–C7–C8 are practically independent of the chosen charge method and scaling factor: left panel (CHelpG), middle panel (RESP-HF), right panel (RESP-B3LYP).

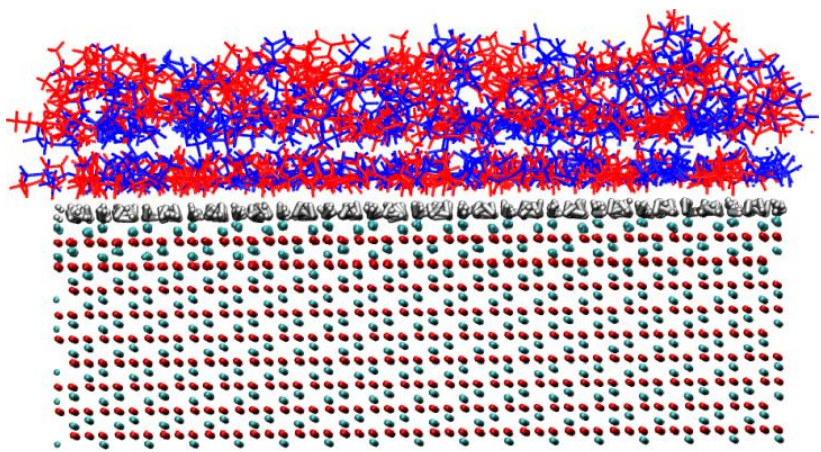

Figure 8. From a snapshot of the MD simulations, a low density region corresponding to the minima in INND (see Fig. 2c) between first few layers and the rest of the ionic liquid is clearly visible.

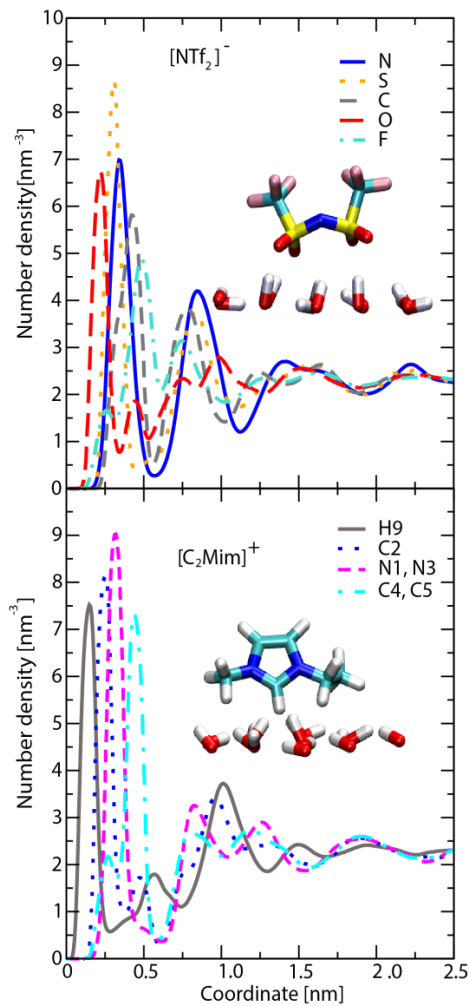

Figure 9. Interface-normal number density per atom type at solid-liquid interface for anion (upper panel) and cation (lower panel). Schematic representation of anion and cation at hydroxylated sapphire surface (small red and white rods).

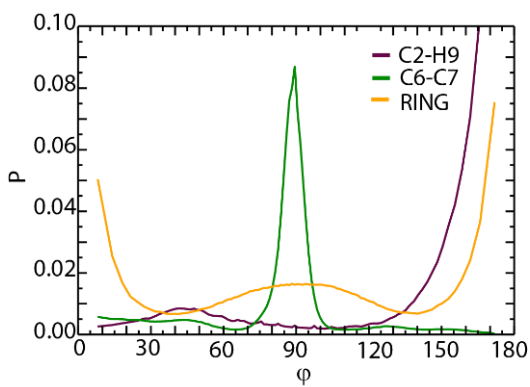

Figure 10. A probability distribution analysis shows the orientation of the C2-H9 bond (maroon line), the alignment of the vector connecting the side carbons C6-C7 (green line) and the orientation of the ring as a whole (orange line), all normal to the interface. The cation ring is mostly perpendicular to the sapphire surface; the short alkyl chains are mostly parallel to the sapphire interface and the hydrogen (H9) points predominantly towards the solid surface.

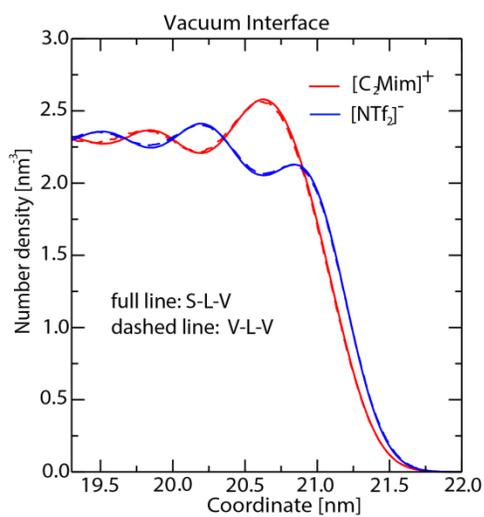

Figure 11. Number density per molecule (red - cation, blue - anion) as a function of z close to the vacuum interface with the sampling time taken to be 160 and 100 ns for S-L-V (full lines) and V-L-V systems (dashed lines), respectively.

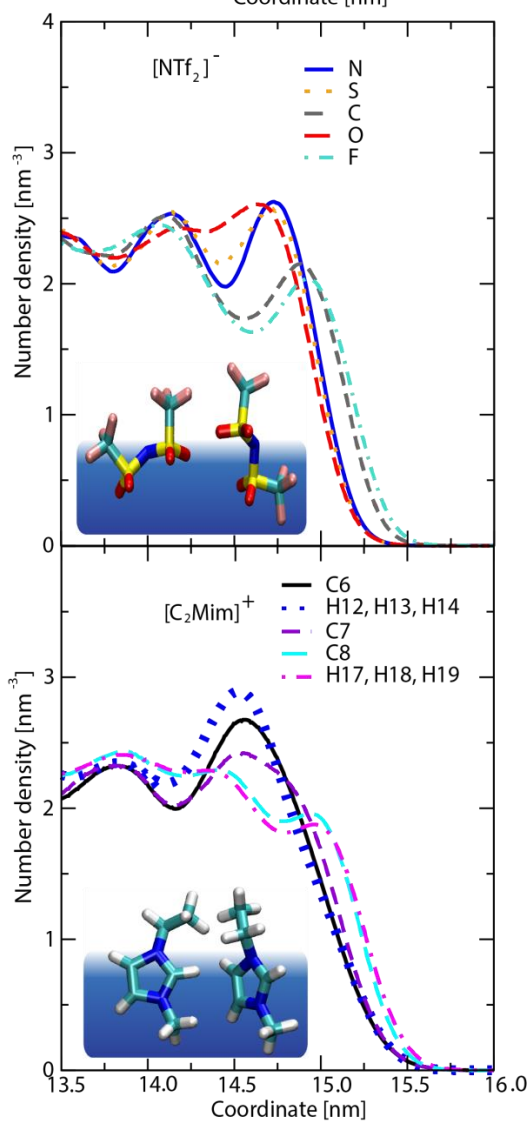

Figure 12. Interface-normal number density per atom type at the liquid-vacuum interface for anion (upper panel) and cation (lower panel). Schematic representation of probable anion and cation conformation at the vacuum interface (dark blue) is also provided.

**Design and Methods**

For all three simulated model systems (Fig. 1b-d), three different parametrization schemes are used for the force field with different charge methods: CHelpG charges,[8] RESP-HF charges[9] (HF/6–31G(d) level of theory) and RESP-B3LYP[10] charges (B3LYP/cc-pVTZ level of theory). In all three parametrization schemes, the Maginn parameters[11] were used for the cation, the CL&P[12] were used for the anion, and the Lorentz-Berthelot mixing rule was applied. The original atomic charges were rescaled to 90% and 85% of their initial values, leading to a total of nine different sets of parameters. The sapphire was optimized in GULP[13] with a fully hydroxylated (0001) x-y surface, described by the CLAYFF[14] force field. Each system was first minimized as per the requirements of the configuration, then relaxed and equilibrated via NVT and NPT ensemble, described in detail in Ref. 1. All simulations for the three model systems were performed in GROMACS 5.1.2[15] with a time step of 2 fs and a cut-off of 2 nm for the van der Waals and short-range Coulomb interactions. Three-dimensional periodic boundary conditions were employed, along with the particle-mesh Ewald procedure for a proper description of the long-range Coulomb interactions for all three model systems. Full details regarding the simulations can be found in Ref. 1.

The bulk liquid (Fig. 1b) was simulated with 1000 ion pairs and had a mix of both *cis* and *trans* configuration for the anions as starting configurations. The V-L-V system was minimized first with 1000 and then with 1400 ion pairs. The introduction of two vacuum interfaces of about 13 nm each to the bulk system of about 8 nm (Fig. 1c) showed that 1400 ion pairs are needed (Fig. 2a-b) for obtaining sufficient homogenized bulk region of about 4 nm which is approximately half the thickness of the IL layer in the V-L-V system. Similar considerations resulted in a system size of 1800 ion pairs of the IL when simulating the S-L-V model system with a column of 16 nm of liquid and a large vacuum (Fig. 1c). This gave a sufficiently large homogenous bulk region as seen in the INND plot of about 7 nm (Fig. 2c). Further, the above mentioned fully homogeneous bulk region is achieved only when the sampling (averaging) time corresponds to the last 70 ns, 100 ns and 160 ns for the L, V-L-V and S-L-V systems, respectively. Details of these criteria for defining a proper bulk region are found in Ref. 1

The large height of the vacuum in the S-L-V system (cf. Fig. 1d) is used to minimize the contribution of the z-replicas (from the periodic boundary conditions imposed) to the electrostatic interactions in the central simulation cell,[16] and to obtain an estimation comparable to that resulting from Ewald sum in the slab geometry. [17,18]

RESP-HF/0.9 was found to be the optimal force field in reproducing various properties of the IL in bulk, in V-L-V and S-L-V system. Details are presented in Ref. 1 and also in Fig. S2 of Supporting Information to Ref. 1.

The depletion region found in Fig. 2 allows for less restricted internal rotations of the anion, intrinsic to its liquid state and is shown in Fig. 5 and discussed in detail in Ref. 1.

## Acknowledgments

We acknowledge funding by the German Research Council, which supports the Excellence Cluster "Engineering of Advanced Materials" at the FAU, support by the DAAD project Multiscale Modelling of Supported Ionic Liquid Phase Catalysis (2017–2018), and the NIC project 11311 at the Jülich supercomputing facilities. R.D.B., C.R.W., A.-S.S. and D.M.S. gratefully acknowledge financial support from the Croatian Science Foundation project CompSoLS-MolFlex (IP-11-2013-8238). We thank Zlatko Brkljača (RBI) for assistance and helpful discussions in the early stages of the project.